\documentstyle[aps,preprint]{revtex}
\begin{document}
\title{Mean Field Approach to Quantum Duffing Oscillator}

\author{Sang Pyo Kim\footnote{Electronic
mail: sangkim@knusun1.kunsan.ac.kr}}

\address{Department of Physics\\
Kunsan National University\\
Kunsan 573-701, Korea}

\maketitle
\begin{abstract}
We propose a mean-field approach to the Duffing oscillator to construct
perturbatively the bounded operators that generate the quantum states,
and to define a non-Gaussian density operator.

\bigskip

PACS number(s): 03.65.-w;  02.30.Mv; 11.80.Fv
\end{abstract}

\bigskip
\bigskip
\bigskip
\bigskip
\bigskip

\centerline{\it Submitted to Letters to Editor of
Journal of Korean Physical Society}
\newpage

The Duffing oscillator is a typical anharmonic oscillator
whose classical aspect is relatively well-understood \cite{nayfeh}.
It has been studied in many fields of physics from classical mechanics
to quantum mechanics, especially as a toy model for one-dimensional
quantum field theory. In particular, more physical interests
have been focused on the application of various perturbation techniques
both as a classical and as a quantum model \cite{bender}.

In this paper we study the quantum Duffing oscillator,
a time-independent quantum anharmonic oscillator,
\begin{equation}
\hat{H} = \frac{\hat{p}^2}{2m} + \frac{m \omega^2 \hat{q}^2}{2}
+ \frac{m \lambda \hat{q}^4}{4}.
\end{equation}
Even though one can find in principle the exact quantum states by solving the
time-independent Schr\"{o}dinger equation, the results of most of attempts
are minimal.

In this Letters to Editor we shall propose a mean-field approach to
the quantum Duffing oscillator, which has the following points different from
other approaches: It takes into account a physically
meaningful frequency that is quite close to the exact classical one.
There appear no secular terms at least at the order of $m \lambda$
for the coupling constant up to a critical value.
A kind of renormalization can be used at the order of $(m\lambda)^2$.
The motivation for this approach is based on two observations:
both the extremization of the Dirac action
and the quantum Liouville equation at the quadratic order
with respect to the vacuum state of a Fock space
lead to the same mean-field equation. We find a complex solution
whose frequency is the same as the exact classical one at the order of
$m \lambda$, construct the operators on the Fock space that generate
the exact quantum states perturbatively, and define a non-Gaussian
density operator.

A general Fock space is constructed by the basis $(\hbar =1)$
\begin{equation}
\hat{a}^{\dagger} = u \hat{p} + v \hat{q},
\hat{a} = u^* \hat{p} + v^* \hat{q}
\end{equation}
where $( u^* v - uv^* ) = i$ follows
from the usual commutation relation $ \left[ \hat{a},
\hat{a}^{\dagger} \right] =1$.
$u$ and $v$ depend on time explicitly, as will shown later.
We extremize the Dirac action \cite{balian}
\begin{equation}
{\cal I} = \int dt \left< \Psi, t \right| i \frac{\partial}{\partial t}
- \hat{H} \left| \Psi, t \right>.
\label{action}
\end{equation}
The variation of Eq. (\ref{action}) with respect to $\left| \Psi, t \right>$
leads to the time-dependent Schr\"{o}dinger equation,
$i \frac{\partial}{\partial t} \left| \Psi, t \right>
= \hat{H} \left| \Psi, t \right>$. The exact quantum state of the form
\begin{equation}
\left| \Psi, t \right> = e^{i \int dt \left< n, t \right| i
\frac{\partial}{\partial t} - \hat{H} \left|n, t \right>} \left| n, t \right>
\end{equation}
gives the null action. As an approximate quantum state we use
the time-dependent number state $ \hat{a}^{\dagger}\hat{a}
\left| n, t \right> = n \left|n, t \right>$ of the Fock space.
Let $v = - m \dot{u}$, $\pi^* = m \dot{u}$, and $\pi = m \dot{u}^*$.
Then, the mean energy of the ground state
\begin{equation}
\left<0,t \right| \hat{H} \left| 0,t \right> =
\frac{\pi^* \pi}{2m} + \frac{m \omega^2}{2} u^* u + \frac{3 \lambda}{4}
(u^* u) ^2,
\end{equation}
acts as a Hamiltonian. The Hamiltonian equations is equivalent
to a second order mean-field equation
\begin{equation}
\ddot{u} + \omega^2 u + 3 \lambda (u^* u) u  = 0.
\label{mean eq}
\end{equation}

We may derive Eq. (\ref{mean eq}) from a different argument,
the quantum Liouville equation
\begin{equation}
i \frac{\partial}{\partial t} \hat{I} + \left[ \hat{I}, \hat{H} \right] = 0.
\label{liouville}
\end{equation}
We expand the Hamiltonian operator in terms of the creation and annihilation
operators
\begin{equation}
\hat{H} = \hat{H}_2 + m \lambda \hat{H}_4 + \left<0,t \right|
\hat{H} \left| 0,t \right>
\end{equation}
where $\hat{H}_2$ and $\hat{H}_4$ are
quadratic and quartic in $\hat{a}^{\dagger}$ and $\hat{a}$.
Then, $\hat{a}^{\dagger}$ and $\hat{a}$ satisfy the approximate
quantum Liouville equation
\begin{equation}
i \frac{\partial}{\partial t} \hat{a}^{\dagger}
+ \left[\hat{a}^{\dagger}, \hat{H}_2 \right]
= 0,
i \frac{\partial}{\partial t} \hat{a}
+ \left[\hat{a}, \hat{H}_2 \right]
= 0.
\label{lowest}
\end{equation}

A complex solution to Eq. (\ref{mean eq})  is found
\begin{equation}
u = \frac{1}{\sqrt{2 m \Omega}} e^{ - i \Omega t},
\label{sol}
\end{equation}
where $\Omega$ satisfies a cubic equation
\begin{equation}
\Omega^3 - \omega^2 \Omega - \frac{3 \lambda}{2m} = 0.
\end{equation}
There is at least one real positive root
\begin{eqnarray}
\Omega = &&  \frac{ 2 ~2^{1/3} \omega^2}{ \Bigl( 324 \frac{\lambda}{m}
+ \sqrt{ - 6912 \omega^6 + 104976 \bigl( \frac{\lambda}{m} \bigr)^2}
\Bigr)^{1/3}}
\nonumber\\
&+& \frac{\Bigl( 324 \frac{\lambda}{m}
+ \sqrt{ - 6912 \omega^6 + 104976 \bigl( \frac{\lambda}{m} \bigr)^2}
\Bigr)^{1/3}}{ 6 ~2^{1/3}}.
\end{eqnarray}
It is remarkable that the mean field equation takes already
the same frequency at the order of $m \lambda$
as the exact classical frequency
\cite{nayfeh}, which is different from those
in Refs. \cite{bender,egusquiza}.

We construct the operators of the quantum Liouville equation (\ref{liouville})
in a perturbative way
\begin{equation}
\hat{A}^{\dagger} = \hat{a}^{\dagger} + \sum_{n = 1}^{\infty}
 \bigl( m \lambda \bigr)^n
\hat{B}_{2n +1}^{\dagger},
\hat{A} = \hat{a}  + \sum_{n = 1}^{\infty}
 \bigl( m \lambda \bigr)^n
\hat{B}_{2n +1}
\label{quant gen}
\end{equation}
where
\begin{equation}
\hat{B}_{2n +1} = \sum_{k = 0}^{2n +1} b^{2n+1}_{k}
\hat{a}^{2n +1 - k} \hat{a}^k.
\end{equation}
For a weak coupling constant we solve Eq. (\ref{liouville})
order by order in $m \lambda$.
The lowest order equation is Eq. (\ref{lowest}), which
has already been solved. At the next order, we obtain
\begin{equation}
\frac{\partial}{\partial t} \hat{B}_3 = i \Bigl(
\left[ \hat{B}_3, \hat{H}_2 \right]
+  \left[ \hat{a}, \hat{H}_4 \right]
+ m \lambda \left[ \hat{B}_3, \hat{H}_4 \right] \Bigr)
\end{equation}
The first bracket on the right hand side is satisfied provided that
$\hat{a}^{\dagger}$ and $\hat{a}$
safisfy Eq. (\ref{lowest}) and all the derivatives act only on the coefficient
functions $b^{3}_k$ but not on $\hat{a}^{\dagger}$ and $\hat{a}$.
The second and third brackets lead to the inhomogeneous equation
\begin{equation}
\frac{d}{dt} \pmatrix{ b^3_0 \cr b^3_1 \cr b^3_2 \cr b^3_3 \cr} =
i m \lambda \pmatrix{ - 9 (u^* u)^2& 0& 3 u^{*4}& 0 \cr
 18 u^* u^3& -3 (u^* u)^2& - 6 u^{*3} u& 9 u^{*4} \cr
- 9 u^4& 6 u^* u^3& 3 (u^* u)^2& - 18 u^{*3} u \cr
0& - 3 u^4 & 0& - 6 (u^* u)^2 \cr}
\pmatrix{ b^3_0 \cr b^3_1 \cr b^3_2 \cr b^3_3 \cr} +
 i \pmatrix{u^{*4} \cr - u^{*3} u \cr (u^* u)^2 \cr - u^* u^3 \cr}.
\label{inh eq}
\end{equation}
For a solution of the form
\begin{equation}
\vec{b}^{3} := (b^3_0, b^3_1, b^3_2, b^3_3)
= (c^3_0 e^{4 i \Omega t},
c^3_1 e^{2 i \Omega t}, c^3_2, c^3_3 e^{- 2i \Omega t} ),
\end{equation}
the homogeneous equation becomes a system of differential equations with
constant coefficients
\begin{equation}
\frac{d}{dt} \pmatrix{c^3_0 \cr c^3_1 \cr c^3_2 \cr c^3_3 \cr} =
i \Omega \pmatrix{ - 9 \alpha - 4 & 0& 3 \alpha & 0 \cr
18 \alpha & - 3 \alpha - 2 & - 6 \alpha & 9 \alpha \cr
- 9 \alpha & 6 \alpha & 3 \alpha & - 18 \alpha \cr
0 & - 3 \alpha & 0 & - 6 \alpha + 2 \alpha \cr}
\pmatrix{c^3_0 \cr c^3_1 \cr c^3_2\cr c^3_3 \cr},
\label{hom eq}
\end{equation}
where $\alpha = \frac{\lambda}{4m \Omega^3}$.
All the eigenvalues of Eq.  (\ref{hom eq}) are real constants
up to the critical coupling constant $\alpha_c = \frac{\lambda_c}{4 m
\Omega^3} \simeq 0.1365$. For the coupling constant larger than
$\lambda_c$, a secular term of an exponential function begins to appear.
This implies that for the strong coupling limit
the Fock space basis may not be
suitable for constructing bounded operators such as $\hat{A}^\dagger$ and
$\hat{A}$ that generate the quantum states. On the other hand,
for the weak coupling
limit ($\alpha << 1$), numericals result show that
$\vec{b}^{3} \sim (c_0^3 e^{-i 3^2 \alpha \Omega t},
c_1^3 e^{-i 3 \alpha \Omega t},
c_2^3 e^{i 3 \alpha \Omega t}, c_3^3 e^{-i 3 \alpha \Omega t})$.
For $\alpha = 0$, the result is exact with the eigevalues
$(-4 \Omega, -2 \Omega, 0, 2\Omega)$.
The freedom in choosing the norms of eigenvectors of Eq. (\ref{hom eq})
may be used to keep the same equation as Eq. (\ref{lowest}), a kind
of renormalization, even at the order of $(m \lambda)^2$.
Thus, the operators, Eq. (\ref{quant gen}), are bounded operators
whose coefficients are proportional to $ \frac{\lambda}{m \Omega^2}$.

Finally, we discuss on the density operator defined as
\begin{equation}
\hat{\rho} = e^{ - \Omega_0 \hat{A}^{\dagger} \hat{A}}.
\label{den op}
\end{equation}
The density operator, Eq. (\ref{den op}), satisfies manifestly
the quantum Liouville equation (\ref{liouville}) by its construction
at the order of $m \lambda$. It would be worthy to
compare it with the Gaussian-type density operator that was
constructed for a time-dependent Duffing oscillator \cite{eboli},
because the operators $\hat{A}^{\dagger}$ and $\hat{A}$
can also work for the time-dependent Duffing oscillator (in fact,
the complex solution (\ref{sol}) is an adiabatic one
in this case) and, furthermore,
Eq. (\ref{den op}) already went beyond the quadratic order
for the Gaussian-type.

The details of this approach and some problems such as
the divergence structure of the operators $\hat{A}^{\dagger}$ and $\hat{A}$
and the relation with the renormalization group approach
\cite{chen,egusquiza}
will be addressed in a future publication \cite{kim2}.

\section*{acknowledgments}
The author would like to thank Daniel Boyanovsky for useful discussions.
This work was supported by Korea Science and Engineering Foundation under
Grant No. 951-0207-056-2.

\end{document}